\documentclass[10pt]{article}

\usepackage{amsmath}
\usepackage{array}
\usepackage{appendix}
\usepackage{graphicx}
\usepackage{amsfonts}
\usepackage{amssymb}
\usepackage{mathrsfs}
\usepackage{yfonts}
\usepackage{euscript}
\usepackage{upgreek}
\usepackage{slantsc}
\usepackage{calligra}
\usepackage[T1]{fontenc}
\usepackage{epsf}
\usepackage{latexsym}

\usepackage{tipa}

\def\be{\begin{equation}}
\def\ee{\end{equation}}
\def\beq{\begin{equation}}
\def\eeq{\end{equation}}
\def\bea{\begin{eqnarray}}
\def\eea{\end{eqnarray}}

\def\ni{\noindent}

\def\!{\hspace{-1.6667em}}

\def\me{\mbox{e}}

\def\bupSigma{\mbox{\boldmath$\Sigma$}}                 % Spatial hypersurface
\def\scE{\mbox{\scriptsize ${\cal E}$}}          % mechanical energy constraint

\def\scL{\mbox{\scriptsize ${\cal L}$}}          % zero total angular momentum constraint
\def\scM{\mbox{\scriptsize ${\cal M}$}}          % momentum constraint of GR  

\def\FrQ{\mbox{\Large $\mathfrak{q}$}}
\def\FrC{\mbox{\Large $\mathfrak{c}$}}

\def\FrS{\mbox{\Large $\mathfrak{s}$}}

\def\FrG{\mbox{\Large $\mathfrak{g}$}}  
\def\FA{\mbox{\Large $\mathfrak{a}$}}

\def\sd{\mbox{\scriptsize d}}
\def\se{\mbox{\scriptsize e}}

\def\sh{\mbox{\scriptsize h}} 
\def\si{\mbox{\scriptsize i}}

\def\sll{\mbox{\scriptsize l}}  %NB EXCEPTIONAL DEF as \sl is reserved for slant.
\def\sm{\mbox{\scriptsize m}}
 
\def\so{\mbox{\scriptsize o}} 
\def\sp{\mbox{\scriptsize p}}

\def\sr{\mbox{\scriptsize r}}
\def\sss{\mbox{\scriptsize s}}  %TO AVOID ARXIV changing \ss to German double s.

\def\sB{\mbox{\scriptsize B}}

\def\sJ{\mbox{\scriptsize J}}

\def\sfA{\mbox{\sffamily{\scriptsize A}}}      % index for general canonical coordinates
\def\pa{\partial}
\def\d{\textrm{d}}

\def\5Star{\mbox{\Large$\star$}}              % big five-point star for hanging things on
\def\sumi2{\sum\mbox{}_{\mbox{}_{\mbox{\scriptsize $i$=1}}}^2}
\def\sumi3{\sum\mbox{}_{\mbox{}_{\mbox{\scriptsize $i$=1}}}^3}

\def\sumIN{\sum\mbox{}_{\mbox{}_{\mbox{\scriptsize $I$=1}}}^{N}}

\def\sumj3{\sum\mbox{}_{\mbox{}_{\mbox{\scriptsize $j$=1}}}^3}
\def\sumk3{\sum\mbox{}_{\mbox{}_{\mbox{\scriptsize $k$=1}}}^3}

\def\timesIN{\bigtimes\mbox{}_{\mbox{}_{\mbox{\scriptsize $I$=1}}}^{N}}
\def\timesI3{\bigtimes\mbox{}_{\mbox{}_{\mbox{\scriptsize $I$=1}}}^{3}}
\def\timesAn{\bigtimes\mbox{}_{\mbox{}_{\mbox{\scriptsize $A$=1}}}^{n}}
\def\bigtimes{\mbox{\Large $\times$}}
\begin{document}

\begin{center}

{\bf \LARGE SPHERICAL RELATIONALISM}

\vspace{.15in}

{\large \bf Edward Anderson} 

\vspace{.15in}

\large {\em DAMTP, Centre for Mathematical Sciences, Wilberforce Road, Cambridge CB3 OWA.  } \normalsize

\end{center}

\begin{abstract}

This paper considers passing from the usual $\mathbb{R}^d$ model of absolute space to $\mathbb{S}^d$ at the level of relational particle models.   
Both approaches' $d = 1$ cases are rather simpler than their $d \geq 2$ cases, 
with $N$ particles in $\mathbb{S}^1$ admitting a straightforward reduction with shape space $\mathbb{T}^{N - 1}$.
The $\mathbb{S}^2$ and $\mathbb{S}^3$ cases -- observed skies and the simplest closed GR cosmologies respectively -- are also considered, 
the latter in the contexts of both static and dynamical radius of the model universe. 
The space of relational triangles on $\mathbb{S}^2$ is hyperbolic 3-space $\mathbb{H}^3$.
Overall, by passing to a closed underlying absolute space, and then to dynamical notion of space, 
we close some of the modelling gaps between relational particle models and geometrodynamics or its inhomogeneous perturbative regime of interest in cosmology.  
Quantum counterparts are also outlined, for use as model arenas of quantum cosmology.
These models are useful in further considerations of both classical and quantum background independence.
 
\end{abstract}

%=========================================================================================================================================================================================
%=========================================================================================================================================================================================
\section{Introduction}\label{Introduction}
%=========================================================================================================================================================================================
%=========================================================================================================================================================================================

Newtonian Mechanics, and the Newtonian Paradigm for Physics more generally, are based on absolute space and time. 
The immovable external character of these led to these being opposed by {\it relationalists}, in particular Leibniz \cite{L} and Mach \cite{M}.  
On the other hand, the Newtonian paradigm of Physics sufficed to explain the observations of nature until the end of the 19th century. 
Furthermore, a satisfactory relational alternative to the foundations of Mechanics was not found until 1982 by Barbour and Bertotti \cite{BB82}.  
Of course, this long post-dates 1) the advent of GR and 2) its dynamical reformulation \cite{ADM-MTW}, by which incorporated a number of relational features into Physics: 
closed universes, dynamical rather than background geometry.
Yet, it turns out that {\it relational particle mechanics (RPMs)} \cite{BB82, B03, FORD, FileR, AMech} are useful model arenas as regards understanding 
many a feature of General Relativity (GR) as a dynamical system \cite{ADM-MTW, FileR}, of GR's configuration spaces \cite{GR-Config, AConfig, AMech}. 
And of analyzing which aspects of Background Independence GR possesses, 
difficulties with which then become facets of the notorious Problem of Time in Quantum Gravity \cite{RPM-PoT, RWR-AM13, ASoS}.  
Indeed, GR admits a geometrodynamical formulation \cite{ADM-MTW} that can be cast in Machian form \cite{RWR-AM13}.
This in the sense of obeying postulates of Temporal and Configurational Relationalism (Sec 2) which can already be set up for RPMs.  
Thus the study of RPMs turns out to have further foundational value (also the above two postulates are also a subset of those of Background Independence \cite{RPM-PoT}).  

\mbox{ }

\ni Each of \cite{AMech} and the current Article extend the range of such relational models in different ways. 
This is to better cover a number of the significant gaps between RPMs and GR. 
The present paper addresses this by use of curved geometry, of closed-universe models, and of dynamical geometry: a range of further models lying in between the two. 
Firstly (Sec 3), I consider relational theory of point particles where $\mathbb{S}^d$ plays the role of absolute space in place of the habitual $\mathbb{R}^d$ of Mechanics.  
These could be the very simple $\mathbb{S}^1$ \cite{FileR}, $\mathbb{S}^2$ either as a model or as the observed sky \cite{Kendall89, Kendall, FileR}, 
or the $\mathbb{S}^3$ that is one of the simplest closed models of space in GR cosmology. 
In Sec 4, I consider the geometry of the spaces of points on a circle and of spherical triangles, alongside Shape Statistics thereupon \cite{Kendall89, Kendall}. 
Shape Statistics of point configurations on the sky is of value to perturbatively inhomogeneous Cosmology \cite{HallHaw, AKendall, AConfig}, 
and means of probing further tests, concepts with interpreting the CMB and the distribution of galaxies. 
Secondly, I consider relational theory on expanding spheres (Sec 5).  
I end by outlining quantum version of spherical RPM in the sense of Dirac quantization in Sec 6.

%=========================================================================================================================================================================================
%=========================================================================================================================================================================================
\section{Outline of the Relational Approach to Physics}
%=========================================================================================================================================================================================
%=========================================================================================================================================================================================

The first structural element in the approach is configuration space $\FrQ$ \cite{Lanczos}: the space of generalized configurations $Q^{\sfA}$ for a physical system.
This paper considers the case of point configurations.  
Here, an incipient $\FrQ$ is 
\be
\FrQ(N, d) = \timesIN \FA(d) \mbox{ } 
\ee 
for $\FA(d)$ the corresponding notion of absolute space. 

\mbox{ } 

\ni Configurational Relationalism is then that 

\mbox{ } 

\ni i)  no extraneous configurational structures 
-- spatial or internal-spatial metric geometry variables that are fixed-background rather than dynamical -- are to be included in the formulation.

\mbox{ } 

\ni ii) Physical formulations in general involve not only a $\FrQ$ but also a $\FrG$ of transformations acting upon $\FrQ$ that are taken to be physically redundant.

% \mbox{ } 

\ni For point particle models, ii) alone is realized [i) is also realized for GR, where the configuration space metric is a function of the 3-metric configuration].  
In some cases, $\FrG$ can be implemented directly, due to enough $\FrG$-invariant objects being available.
More widely, $\FrG$ can be implemented indirectly, whether by A) Best Matching \cite{BB82}, 
which involves $\FrG$ corrections to the theory's changes of redundant configuration, $\d q_I$ (see Sec 3 for examples).  
Or B) by the more general `$\FrG$-act $\FrG$-all' method (see e.g. \cite{AMech}).

\mbox{ } 

\ni On the other hand, Temporal Relationalism -- the Leibnizian stance that that there is no time at the primary level for the universe as a whole -- can be formulated as follows.

\mbox{ } 

\ni i)  The action is not to include any extraneous times -- such as Newtonian time -- or extraneous time-like variables, 
such as the lapse function \cite{ADM-MTW} in the geometrodynamical formulation of GR.

\mbox{ } 

\ni ii) Time is not to be smuggled into the action in the guise of a label either.

\mbox{ } 

\ni Here, ii) is most optimally implemented by a geometrical action, which is dual to parametrization irrelevant action, 
which upon parametrization would become a reparametrization invariant action \cite{FileR, TRiPoD}. 
The particular cases of geometrical actions in the present Article are all {\it Jacobi-type} action \cite{Lanczos}, 
\beq
S = \sqrt{2}\int \d s \sqrt{E - V} \mbox{ } , \mbox{ } \mbox{ } 
\label{S-rel}
\eeq
corresponding to a Riemannian notion of geometry.
A primary constraint must then follow from these actions due to an insight of Dirac's \cite{Dirac}.  
In the case of Mechanics, this gives an equation which is usually interpreted as an energy conservation equation. 
However in the present context this is to be reinterpreted as an {\it equation of time} \cite{B94I}. 
Indeed rearranging it gives an expression for emergent Machian time: a concrete realization resolution of primary-level timelessness by Mach's `time is to be abstracted from change'.
In the case of metric Scale-and-Shape Mechanics, this emergent time amounts to a relational recovery of a quantity which is more usually regarded as Newtonian time.
Note that more generally handling Mechanics models in a temporally relational manner requires a modified version of the Principles of Dynamics as laid out in \cite{TRiPoD}.

\mbox{ } 

\ni Another useful notion is {\it relational nontriviality}, which requires at least 2 degrees of freedom, so that the value of one can be expressed in terms of the other 
(rather than in terms of a no longer a priori existent external time parameter).
This leads to the notion of a {\it smallest relationally nontrivial unit} for each relational theory \cite{AMech}. 
This is concurrently the smallest relational universe model, the smallest relationally nontrivial subsystem, and the smallest Shape Statistics sampling probe.
An archetype for this is the relational triangle in 2- and 3-$d$ metric scale-and-shape and pure-shape RPMs, 
as features in both Barbour's familiar demonstration of Best Matching with two wooden triangles and in Kendall's sampling by triples of points in 2-$d$. 
Furthermore, the geometry of the shape space of these probes is geometrically well-understood: it is $\mathbb{S}^2$. 
(Or a piece thereof: Kendall's spherical blackboard \cite{Kendall84, Kendall89}.  
See \cite{AConfig} for discussion of various pieces of this that are relevant in different modelling contexts.)
This $\mathbb{S}^2$ geometry then enters many subsequent calculations for the corresponding Shape Mechanics and Shape Statistics \cite{Kendall, FileR, AKendall}.

%=========================================================================================================================================================================================
%=========================================================================================================================================================================================
\section{RPMs for particles on spheres}\label{Sphe-RPM}
%=========================================================================================================================================================================================
%=========================================================================================================================================================================================

%=========================================================================================================================================================================================
\subsection{General $\mathbb{S}^p$ case}
%=========================================================================================================================================================================================

Here absolute space is modelled by $\FA = \mathbb{S}^p$, 
\be
\FrQ(N, d) = \timesIN \mathbb{S}^p \mbox{ } ,
\ee 
and a natural and nontrivial choice of $\FrG$ is 
\be 
Isom(\mathbb{S}^p) = Rot(p+ 1) = SO(p + 1) \mbox{ } . 
\ee
The relational action for this is built out of a potential function $V = V(\mbox{$p$--}\sphericalangle)$ -- i.e. a function of relative $p$-sphere angles alone -- 
and a curvilinear rendition of Best Matching based upon $SO(p + 1)$ auxiliaries $S_{[ij]}$: (\ref{S-rel}) with 
\beq
\d s^2 = \sumIN \d_Ss^2 \mbox{ } . 
\label{S-sphe-p}
\eeq
Then the quadratic constraint
\beq
\scE := \sumIN N^{pIqJ}p_{pI}p_{qJ}/2 + V(\mbox{$p$--}\sphericalangle) = E
\label{E-sphe-p}
\eeq
arises as a primary constraint,           for 
\be
N^{pIqJ} := \prod\mbox{}_{\mbox{}_{\mbox{\scriptsize $m$ = 1}}}^{p - 1} \mbox{sin}^{-2}\theta_{mI} \delta^{pq}\delta^{IJ}
\ee 
the inverse of the standard $p$-sphere metric,  
\be
M_{pIqJ} := \prod\mbox{}_{\mbox{}_{\mbox{\scriptsize $m$ = 1}}}^{p - 1} \mbox{sin}^2   \theta_{mI} \delta_{pq}\delta_{IJ} \mbox{ } .
\ee 
($\theta^{mI}$ here are $N$ copies of the  $p$-sphere angle coordinates and $E$ is total energy of the model universe.)

\mbox{ } 

\ni Also the linear constraint $\scM_{ij} = 0$ arises as a secondary constraint from variation with respect to the $S_{[ij]}$.  

\mbox{ }

\ni Moreover (\ref{E-sphe-p}) can indeed be rearranged to give an expression for classical Machian emergent time: 
\beq
t^{\se\sm(\sJ\sB\sB)} = \mbox{\Large E}_{S \in SO(p + 1)}\left( \int \d s  \mbox{\LARGE /}  \sqrt{2\{E - V(\mbox{$p$--}\sphericalangle)\}} \right) \mbox{ } . 
\eeq
$\d s$ here depends implicitly on the $SO(p +1)$ auxiliaries via (\ref{S-sphe-p}), whereas $\mbox{\Large E}$ denotes `extremization over'.

\ni The minimal relational nontrivial units have $2 \leq \mbox{dim}(\FrQ) - \mbox{dim}(\FrG) = Np - \{p + 1\}p/2 = p\{2N - p - 1\}/2$, 
giving 3 points on   the circle for $p = 1$, 
       the   spherical triangle for $p = 2$,  
   and the 3-spherical triangle for $p = 3$.

%=========================================================================================================================================================================================
\subsection{Specifics of $\mathbb{S}^1$ `circle lines'}
%=========================================================================================================================================================================================

In this case, absolute space is modelled by $\FA = \mathbb{S}^1$, 
\be
\FrQ = \timesIN \mathbb{S}^1 \mbox{ } , 
\ee 
and a natural and nontrivial choice of $\FrG$ is 
\be
Isom(\mathbb{S}^1) = Rot(2) = SO(2) = U(1) \mbox{ } . 
\ee
Extending the nomenclature of \cite{AF}, who considered particles on a line as `metro line configurations', the current theories' configurations 
are topologically circular lines, such as indeed the London Underground's circle line. 
The spaces of these configurations then form $N$-stop metroland for the $\mathbb{R}$ case and {\it N-stop circle-line-land} for the new $\mathbb{S}^1$ case.  

\mbox{ }

\ni The relational action for $N$-stop circle-line-land is then built out of $V = V(\mbox{relative angles } \phi^I - \phi^J)$ 
and the Best Matching 
\be
\d \theta_{\alpha}^I := \d\theta^I - \d\alpha \mbox{ } :
\ee
(\ref{S-rel}) with 
\beq
\d s^2 = \sumIN \d_{\alpha}\phi_I^2 \mbox{ } . 
\label{S-circ}
\eeq
The conjugate momenta are then 
\beq
p^{\theta}_I = \frac{\sqrt{2 \{E - V(\phi^I - \phi^J)\}}}{\d s} \d_{\alpha}\phi^I \mbox{ } .
\eeq
These obey 
\beq
\scE := \sumIN p^{\phi \, 2}_I/2 + V(\phi^I - \phi^J) = E
\label{E-sphe}
\eeq
as a primary constraint, and $\scL = 0$ as a secondary constraint from variation with respect to the single $Rot(2)$ auxiliary, $\alpha$.

\mbox{ } 

\ni For these models, it is immediate to solve $\scL = 0$ as an equation for $\alpha$, giving 
\be
\d\alpha = \sumIN\d\phi^I/N := \langle\d\phi\rangle \mbox{ } : 
\ee
an {\it averaged object}. 
Then substituting back into the kinetic arc element, 
\be
\d \widetilde{s}^2 = \sumIN \d\bar{\phi}^{I\,2}
\ee
for {\it contrast objects} defined according to  $\bar{O} := O - \langle O \rangle$.  
There are $n = N - 1$ independent such, amounting to losing one $\mathbb{S}^1$ in the quotienting.  
In some ways, this is the circular analogue of passing to centre of mass frame in quotienting out translations.  

\mbox{ }

\ni Finally note that (\ref{E-sphe}) can be rearranged to give
\beq
t^{\se\sm} = \mbox{\Large E}_{\alpha \in Rot(2)}\left( \int \d s \mbox{\LARGE /} \sqrt{2\{E - V(\phi^I - \phi^J)\}} \right) 
           = \int \d \widetilde{s} \mbox{\LARGE /} \sqrt{2\{E - V(\phi^I - \phi^J)\}} 
\eeq
for $\d s$ depending implicitly on $\alpha$ through (\ref{S-sphe}).
The last more explicit line is obtained through the above reduction.

%FFFFFFFFFFFFFFFFFFFFFFFFFFFFFFFFFFFFFFFFFFFFFFFFFFFFFFFFFFFFFFFFFFFFFFFFFFFFFFFFFFFFFFFFFFFFFFFFFFFFFFFFFFFFFFFFFFFFFFFFFFFFFFFFFFFFFFFFFFFFFFFFFFFFFFFFFFFFFFFFFFFFFFFFFFFFFFFFFFFFFFFFFF
{            \begin{figure}[ht]
\centering
\includegraphics[width=0.4\textwidth]{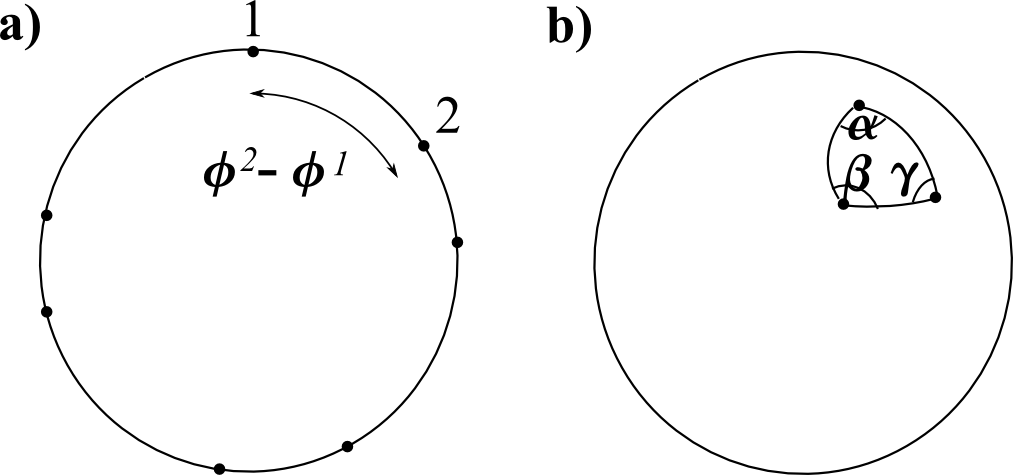}
\caption[Text der im Bilderverzeichnis auftaucht]{        \footnotesize{A circle line configuration, and a minimal relationally nontrivial spherical triangle configuration.} }
\label{Fig-1}\end{figure}            }
%FFFFFFFFFFFFFFFFFFFFFFFFFFFFFFFFFFFFFFFFFFFFFFFFFFFFFFFFFFFFFFFFFFFFFFFFFFFFFFFFFFFFFFFFFFFFFFFFFFFFFFFFFFFFFFFFFFFFFFFFFFFFFFFFFFFFFFFFFFFFFFFFFFFFFFFFFFFFFFFFFFFFFFFFFFFFFFFFFFFFFFFFFF
%
%==========================================================================================================================================================================================
\subsection{Specifics of $\mathbb{S}^2$ `constellations'}
%==========================================================================================================================================================================================

Now, absolute space is modelled by $\FA = \mathbb{S}^2$, 
\be
\FrQ = \timesIN \mathbb{S}^2 \mbox{ } ,
\label{Q-S2}
\ee 
and a natural and nontrivial choice of $\FrG$ is 
\be
Isom(\mathbb{S}^2) = Rot(3) = SO(3) \mbox{ } . 
\ee
It is particularly appropriate then to term the point particle configurations in this case {\it constellations}, the corresponding space of which is {\it constellationland}.

\mbox{ } 

\ni The corresponding rendition of Best Matching corrections now take a curvilinear form:
\be
\d_{\underline{B}}\theta^I := \d \theta^I + \mbox{cos}\,\phi^I \, \d B_2 -  \mbox{sin}\,\phi^I \, \d B_1 \mbox{ } , 
\eeq 
\beq 
\d_{\underline{B}}\phi^I := \d \phi^I + \d B_3 - \mbox{cot}\theta^I\{\mbox{sin}\,\phi^I \, \d B_2 + \mbox{cos}\,\phi^I \, \d B_1\} \mbox{ } .
\eeq
The relational action is (\ref{S-rel}) with $V = V(\sphericalangle)$ alone and 
\beq
\d s^2 = \sumIN \{ \d_{\underline{B}}\theta_I^2 + \mbox{sin}^2\theta_I \d_{\underline{B}}\phi_I^2 \} \mbox{ } .  
\label{S-sphe}
\eeq
Then the conjugate momenta are
\beq
p^{\theta}_I = \frac{\sqrt{2 \{E - V(\sphericalangle)\}}}{\d s} \d_{\underline{B}}\theta^I \mbox{ } , \mbox{ } \mbox{ } 
p^{\phi}_I   = \mbox{sin}^2\theta_I \, \frac{\sqrt{2 \{E - V(\sphericalangle)\}}}{\d s} \d_{\underline{B}}\phi^I  \mbox{ } .
\eeq
These obey 
\beq
\scE := \sumIN\{p^{\theta \, 2}_I + \mbox{sin}^{-2}\theta^I \, p^{\phi \, 2}_I\}/2 + V(\sphericalangle) = E
\label{E-sphe-2}
\eeq
as a primary constraint, and 
\be
\underline{\scL} = \sumIN
\left[ 
- \mbox{sin}\,\phi^I                       p^{\theta}_I 
- \mbox{cos}\,\phi^I \mbox{cot}\,\theta_I  p^{\phi}_I     \mbox{ } , \mbox{ } \mbox{ }					
  \mbox{cos}\,\phi^I                       p^{\theta}_I 
- \mbox{sin}\,\phi^I \mbox{cot}\,\theta_I  p^{\phi}_I     \mbox{ } , \mbox{ } \mbox{ }									
                                           p^{\phi}_I    
\right] = 0                                               \mbox{ } , \mbox{ } \mbox{ }
\ee
as a secondary constraint from variation with respect to $\underline{B}$.

\mbox{ } 

\ni Furthermore, (\ref{E-sphe-2}) can be rearranged to give the emergent Machian time
\beq
t^{\se\sm} = \mbox{\Large E}_{\underline{B} \in SO(3)}\left( \int\d s/\sqrt{2\{E - V(\sphericalangle)\}} \right)
\eeq
for $\d s$ depending implicitly on $\underline{B}$ as per (\ref{S-sphe}).

%==========================================================================================================================================================================================
\subsection{2-$d$ with other metrics}
%===========================================================================================================================================================================================

As a useful digression at this point, the general ellipsoidal line element is 
[e.g. using $\underline{x} = (a \, \mbox{sin}\,\theta\, \mbox{sin}\,\phi, b \, \mbox{sin}\theta\, \mbox{cos}\,\phi, c \, \mbox{cos}\,\theta)$ in a fiducial embedding flat $\mathbb{R}^3$]
$$
\d s^2_{\se\sll\sll\si\sp\sss\so\si\sd} =     \{ \mbox{cos}^{2}\theta \{ a^2 \mbox{sin}^{2}\phi + b^2 \mbox{cos}^{2}\phi \} + c^2 \mbox{sin}^{2}\theta \}  \d \theta^2
$$                                    
\beq									
									+ 2 \{a^2 - b^2\} \mbox{sin}\,\theta \, \mbox{cos}\,\theta \, \mbox{sin}\,\phi \, \mbox{cos}\,\phi  \d \theta \d \phi
			                        +                \mbox{sin}^{2}\theta \{a^2 \, \mbox{cos}^{2}\theta + b^2 \mbox{sin}^{2}\theta\} \d \phi^2     \mbox{ } .
\eeq
Now for the spheroidal special case $a = b$, the line element collapses to 
\beq
\d s^2_{\sss\sp\sh\se\sr\so\si\sd} =  \{ a^2 \, \mbox{cos}^{2}\theta_I + c^2 \mbox{sin}^{2}\theta_I \}  \d \theta^2
			                       +     a^2 \, \mbox{sin}^{2}\theta_I                                  \d \phi^2    \mbox{ } , 
\eeq
so $\phi$ is a cyclic coordinate, corresponding to an axial symmetry $SO(2)$. 
In this case (and for $c \neq a = b$ so as to not fall back to the previous Sec's {\sl very} special case), 
\be
Isom(\langle\mathbb{S}^2, \d s^2_{\sss\sp\sh\se\sr\so\si\sd}\rangle) = SO(2)
\ee 
is available as a $\FrG$ of physically meaningless transformations.
Then take $\FrQ$ as per (\ref{Q-S2}) with each $\mathbb{S}^2$ equipped with this metric. 
This amounts to summing over $N$ labelled copies of this metric, alongside applying the Best Matching correction $\d_{\underline{B}}\phi^I := \d \phi^I + \d B^3$.
This makes sense in conjunction with the $SO(2)$ respecting form  $V = V(\theta^I \mbox{alone})$. 

\mbox{ } 

\ni On the other hand, for the generic ellipsoidal case -- $a, b, c$ all distinct -- 
\be 
Isom(\langle\mathbb{S}^2, \d s^2_{\se\sll\sll\si\sp\sss\so\si\sd}\rangle) = id \mbox{ } ,
\ee 
so no continuous $\FrG$ is available in this case. 
This is indicatory of the generic situation for RPMs on fixed manifolds: most fixed manifolds have enough structure that there is no scope for any Configurational Relationalism at all.
Moreover, this is a limitation of RPM modelling which the GR situation itself does not succumb to.  

\mbox{ } 

\ni The above examples indicate that there can be multiple metrics on a manifold, which exhibit, moreover, a diversity of isometry groups.

%=========================================================================================================================================================================================
\subsection{3-$d$ closed universe particle model}
%=========================================================================================================================================================================================

Here absolute space is modelled by $\FA = \mathbb{S}^3$, 
\be 
\FrQ = \timesIN \mathbb{S}^3 \mbox{ } , 
\ee 
and a natural and nontrivial choice of $\FrG$ is 
\be
Isom(\mathbb{S}^3) = SO(4) \mbox{ } .
\ee 
The relational action is then built out of $V = V(\mbox{3--}\sphericalangle)$ and another use of Best Matching in curvilinear form.
The action here is (\ref{S-rel}) with 
\beq
\d s^2 = \sumIN \{ \d_{S}\theta_I^2 + \mbox{sin}^2\theta^I \d_{S}\phi_I^2 + \mbox{sin}^2\theta^I \mbox{sin}^2\phi^I \d_{S}\varphi_I^2\} \mbox{ } 
\label{S-sphe-2}
\eeq
now involving two 3-vectors' worth of auxiliaries $S_{[ij]}$: $\underline{B}$ and $\underline{B}^{\prime}$, due to the underlying `accidental relation'
\be
SO(4) \cong SO(3) \times SO(3) \mbox{ } .
\ee  
\ni The ensuing constraints are 
\beq
\scE := \sumIN\{p^{\theta \, 2}_I + \mbox{sin}^{-2}\theta^I \, p^{\phi \, 2}_I \mbox{sin}^{-2}\theta^I\mbox{sin}^{-2}\phi^I \, p^{\varphi \, 2}_I\}/2 + V(\mbox{3--}\sphericalangle) = E
\label{E-sphe-3}
\eeq
as a primary constraint, 
and the $SO(4)$ constraint ${\scM}_{ij} = 0$ reducing to the schematic form $\underline{\scL} = 0$ and $\underline{\scL}^{\prime} = 0$, 
as a secondary constraint from variation with respect to $\underline{B}$ and $\underline{B}^{\prime}$.

\mbox{ } 

\ni Also now
\beq
t^{\se\sm} = \mbox{\Large E}_{\underline{B}, \underline{B}^{\prime}, \in SO(3) \times SO(3)}\left( \int\d s/\sqrt{2\{E - V(\sphericalangle)\}} \right)
\eeq
for $\d s$ depending implicitly on $\underline{B}$ and $\underline{B}^{\prime}$ through (\ref{S-sphe-2}).

\mbox{ } 

\ni Finally note that subgroups of $SO(d + 1)$ present further options, and that affine and conformal extensions of Relationalism -- considered in \cite{AMech} for $\mathbb{R}^d$ -- 
are also available for $\mathbb{S}^d$.

%=========================================================================================================================================================================================
%=========================================================================================================================================================================================
\section{Shape space geometry and Shape Statistics}
%=========================================================================================================================================================================================
%=========================================================================================================================================================================================

%FFFFFFFFFFFFFFFFFFFFFFFFFFFFFFFFFFFFFFFFFFFFFFFFFFFFFFFFFFFFFFFFFFFFFFFFFFFFFFFFFFFFFFFFFFFFFFFFFFFFFFFFFFFFFFFFFFFFFFFFFFFFFFFFFFFFFFFFFFFFFFFFFFFFFFFFFFFFFFFFFFFFFFFFFFFFFFFFFFFFFFFFF
{            \begin{figure}[ht]
\centering
\includegraphics[width=0.65\textwidth]{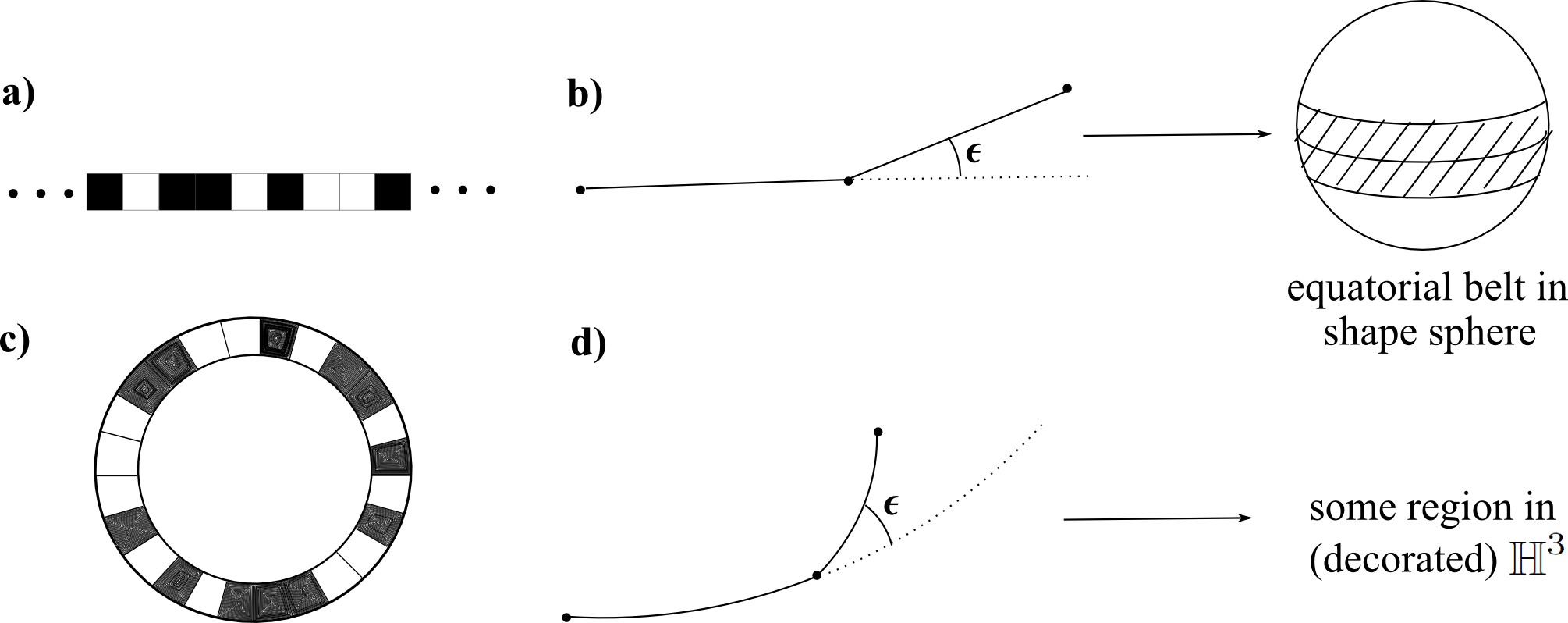}
\caption[Text der im Bilderverzeichnis auftaucht]{        \footnotesize{a) Clumping on a line. 
b) Set-up for angular Shape Statistics in $\mathbb{R}^2$ space. 
c) Clumping on a circle.
d) Set-up for angular Shape Statistics in $\mathbb{S}^2$ space; great sphere geodesics have taken the place of lines, and `decorated' refers to non-principal strata.} }
\label{Fig-2}\end{figure}            }
%FFFFFFFFFFFFFFFFFFFFFFFFFFFFFFFFFFFFFFFFFFFFFFFFFFFFFFFFFFFFFFFFFFFFFFFFFFFFFFFFFFFFFFFFFFFFFFFFFFFFFFFFFFFFFFFFFFFFFFFFFFFFFFFFFFFFFFFFFFFFFFFFFFFFFFFFFFFFFFFFFFFFFFFFFFFFFFFFFFFFFFFFF

The general shape space for $N$ particles on $\mathbb{S}^d$ is 
\beq
\FrC(N, d) := \timesIN \mathbb{S}^d/SO(d + 1) \mbox{ } 
\eeq
In the 1-$d$ case, 
\beq
\mbox{\it circle-line-land } \mbox{ } \FrC(N, 1)    := \timesIN \mathbb{S}^1/SO(2) = \timesAn \mathbb{S}^1 = \mathbb{T}^n   
\eeq
and 
\beq
Isom(\FrC(N, 1)) = Isom(\mathbb{T}^n) = \timesAn U(1) \mbox{ } .
\eeq
\ni On the other hand, in the 2-$d$ case, {\it constellationland} $\FrC(N, 2) := \timesIN \mathbb{S}^2/SO(3)$.
The smallest relationally nontrivial such is {\it spherical triangleland} $\FrC(3, 2) := \timesI3 \mathbb{S}^1/SO(3)$, 
whose principal stratum \cite{AConfig} is of dimension $2 \times 3 - 3 = 3$.
Furthermore, this is geometrically \cite{Kendall} the hyperbolic space
\beq
\mathbb{H}^3 := \{x \in \mathbb{M}^{4}|( ||\underline{x}||^2 = -1, x_0 > 0\} \mbox{ } ,
\eeq 
(here $\mathbb{M}^4$ is mathematically the same as 4-$d$ Minkowski spacetime), with standard line element 
\be
\d s = \frac{2||\d \underline{x}||}{1 - ||\underline{x}||^2} \mbox{ } . 
\ee 
Then 
\beq
Isom(\mathbb{H}^3) =  \mbox{ 2-$d$ M\"{o}bius group } \cong SO^+(3, 1) \mbox{ } :  
\eeq
the proper orthochronous Lorentz group so familiar from SR. 

\mbox{ } 

\ni In contrast, the metric shape space $\FrS(N, d)$ on $\mathbb{R}^d$ with $\FrS(N, 1) = \mathbb{S}^{n - 1}$ and $\FrS(N, 2) = \mathbb{CP}^{n - 1}$ \cite{Kendall, FileR} .  
This shows some capacity for arbitrary-$N$ shape spaces being geometrically standard (though this does not carry over to 3-$d$).  
Some remaining questions then concern the topological and geometrical identity of $\FrC(N, 2)$ for $N > 3$, and of the probe space $\FrC(N, 3)$.
Reduction of the Mechanics action can on occasion point to these features \cite{FORD, FileR}.
Though metric shapes on $\mathbb{S}^2$ parallel metric shapes on $\mathbb{R}^3$ [both involve $SO(3)$] in only having locally attainable reduction, 
and metric shapes on $\mathbb{S}^3$ are also expected to follow suit in this regard.

\mbox{ } 

\ni As regards clumping statistics on circles, Roach's \cite{Roach} black and white square procession approach on $\mathbb{R}^1$ can be extended to this case. 
This extension involves truncating the geometric progression of black-square chain-lengths and averaging over choice of starting-point for the procession on the circle is performed.

\mbox{ } 

\ni As an alternative, one could consider Shape Statistics based on Geometrical Probability on $\FrS(N, 1) = \mathbb{S}^n$: $N$-stop metroland \cite{AF}, 
with circle-line-land counterpart of  Shape Statistics based on Geometrical Probability on $\FrC(N, 1) = \mathbb{T}^n$: $N$-stop circle-line-land.  

\mbox{ }

\ni Shape Statistics on circles additionally involves relative angle information (the circle clumping case's angle is but an extrinsic notion). 
This parallels how $\mathbb{R}^1$ has just clumping (ratio) information whereas $\mathbb{R}^2$ has both ratio and angle information, 
thus necessitating angle-based Shape Statistics like Kendall's \cite{Kendall84, Kendall}.
The $\mathbb{S}^2$ angular Shape Statistics can be use e.g. to investigate quasar alignment in the observed sky \cite{Quasar}. 
More generally, shape space geometry and Shape Statistics provide the means of devising further and mathematically better-founded pattern tests along the lines of `circles in the sky'  
or `multiple image identification' \cite{LL}.

%=========================================================================================================================================================================================
%=========================================================================================================================================================================================
\section{Minisuperspace-RPM coupled models}
%=========================================================================================================================================================================================
%=========================================================================================================================================================================================

One could furthermore entertain the notion of space $\bupSigma$ itself being dynamical: a GR-type feature.  
E.g. $\bupSigma$ could undergo whichever of expansion, anisotropic change, and inhomogeneous change.   
A further source of variety in these cases is whether to model particles on a dynamical $\bupSigma$, fields on a dynamical $\bupSigma$ or just the dynamics of $\bupSigma$ itself. 
Consider for now an isotropic GR model coupled to spherical RPM. 
%
% Isotropic is here much lika an opposite-sign version of scaled RPM cone. 
%
These are in essence a combined minisuperspace--RPM model, with
\be
S = 2\int\sqrt{  \frac{\kappa a \, \d a^2}{4} - a^2\frac{\d s^2_{\mathbb{S}^3}}{2}  }    \sqrt{2\kappa a\{\Lambda a^2 - 3) + V(\mbox{3--}\sphericalangle, a)}
\ee
Here $\kappa$ is the gravitational part of the action's dimensionful prefactor, $a$ is the cosmological scalefactor and $\Lambda$ is the cosmological constant.    

\mbox{ }

\ni The conjugate momenta are then 
\be
\pi_a            = \frac{    \sqrt{  2\kappa a\{\Lambda a^2 - 3) + V(\mbox{3--}\sphericalangle, a)  }    }{\d s} \frac{\kappa a}{2} \d a        \mbox{ } , \mbox{ } \mbox{ } 
p^{\theta}_{I i} = \frac{    \sqrt{  2\kappa a\{\Lambda a^2 - 3) + V(\mbox{3--}\sphericalangle, a)  }    }{\d s}         a^2 M_{iIjJ} \d \theta^{jJ} \mbox{ } .
\ee
These obey the primary constraint 
\be
-\frac{\pi_a^2}{\kappa a} + \frac{||p^{\theta}||_{\times_{I = 1}^N \mathbb{S}^3}}{2a^2}\mbox{ }^2 = 2\kappa a\{3 - \Lambda a^2\} - V(\mbox{3--}\sphericalangle, a) \mbox{ } .  
\ee
\ni Finally, 
\beq
t^{\se\sm} = \int \sqrt{    \frac{  -\kappa \d a^2 + 2a^2\d s^2_{\mathbb{S}^3}  }{  4\{2\kappa a^2\{3 - \lambda a^2\} - V(\mbox{3--}\sphericalangle, a)  }    } \approx  
t^{\se\sm}_0 - \frac{1}{2\sqrt{2}}\int\frac{a}{\Lambda a^2 - 3} \frac{\d s_{\mathbb{S}^3}}{\d a} + 
               \frac{1}{128\sqrt{\kappa}}  \int  \frac{V(\mbox{3--}\sphericalangle)}{a^4\{3 - \Lambda a^2\}^2} \mbox{ } . 
\eeq
(compare the expansions of classical Machian emergent time in \cite{ACos2-AMSS}).  

\mbox{ } 

\ni Note 1) As regards the particles breaking the homogeneity, in any case we want the inhomogeneity to be small for the cosmological applications in hand, 
so the spherical RPM's are to be treated as perturbations about minisuperspace.

\ni Note 2) A limitation in this analogy is that GR's inhomogeneous perturbations about minisuperspace have a further scalar--vector--tensor (SVT) split of modes. 

\ni Note 3) On the other hand, a virtue of this Sec's model arena is the availability of further pattern analysis for it, of use in analysis of cosmological structure formation.
This remains a distant prospect for actual inhomogenous GR.

%=========================================================================================================================================================================================
%=========================================================================================================================================================================================
\section{Quantum spherical RPMs}
%=========================================================================================================================================================================================
%=========================================================================================================================================================================================

\ni Dirac quantization is straightforward.
Here the standard spherical kinematical quantization \cite{I84} will do. 

\mbox{ }

\ni In the 1-$d$ case, this gives
\be
\widehat{\scL}\Psi = \mbox{$\frac{\hbar}{i}$} \sumIN \mbox{$\frac{\pa\Psi}{\pa\phi^I}$} = 0 
\ee
-- meaning that $\Psi = \Psi(\phi^I - \phi^J)$ alone -- and the main wave equation 
\be
\widehat{\scE}\Psi = -\mbox{$\frac{\hbar}{2}$} \sumIN \mbox{$\frac{\pa^2\Psi}{\pa\phi^{I\,2}}$} + V(\phi^I - \phi^J)\Psi = E\Psi \mbox{ } . 
\ee
\ni On the other hand, the 2-$d$ case gives  
\be
\widehat{\underline{\scL}}\Psi = \mbox{$\frac{\hbar}{i}$}\sumIN
\left[ 
- \mbox{sin}\,\phi^I                      \mbox{$\frac{\pa\Psi}{\pa \theta^I}$} 
- \mbox{cos}\,\phi^I \mbox{cot}\,\theta_I \mbox{$\frac{\pa\Psi}{\pa \phi^I}$}     \mbox{ } , \mbox{ } \mbox{ }					
  \mbox{cos}\,\phi^I                      \mbox{$\frac{\pa\Psi}{\pa \theta^I}$} 
- \mbox{sin}\,\phi^I \mbox{cot}\,\theta_I \mbox{$\frac{\pa\Psi}{\pa \phi^I}$}     \mbox{ } , \mbox{ } \mbox{ }									
                                          \mbox{$\frac{\pa\Psi}{\pa \phi^I}$}    
\right] = 0                                                                       \mbox{ } , \mbox{ } \mbox{ }
\ee
meaning that $\Psi = \Psi(\sphericalangle)$ alone, alongside the main wave equation 
\be
\widehat{\scE}\Psi = -\mbox{$\frac{\hbar}{2}$} \sumIN D^2_{\mathbb{S}^2} \Psi + V(\sphericalangle)\Psi = E\Psi \mbox{ } .  
\ee
\ni The 3-$d$ case then necessitates 6 components' worth of
\be
\widehat{{\scM}_{ij}}\Psi = 0 \mbox{ } : \widehat{\underline{\scL}}\Psi = 0 \mbox{ } \mbox{ and } \mbox{ } \widehat{\underline{\scL}^{\prime}}\Psi = 0 \mbox{ } , 
\label{Hat-S-3}
\ee
meaning that $\Psi = \Psi(\mbox{3--}\sphericalangle)$ alone, and the main wave equation 
\be
\widehat{\scE}\Psi = -\mbox{$\frac{\hbar}{2}$} \sumIN D^2_{\mathbb{S}^3} \Psi + V(\mbox{3--}\sphericalangle)\Psi = E\Psi \mbox{ } .  
\ee
Finally the 3-$d$ case coupled to a GR-type dynamical scalefactor gives (\ref{Hat-S-3}) alongside the new main wave equation 
$$
\widehat{\scE}\Psi \mbox{ } = \mbox{ } -\frac{\hbar^2}{2} 
\left\{
-\frac{2}{\kappa a}
\left\{
\frac{\pa^2}{\pa a^2} + \{3N - \mbox{$\frac{1}{2}$}\}\frac{\pa}{\pa a} 
\right\} + 
\right. \hspace{3in}
$$
\be
\left.
\frac{1}{a^2}\sumIN D^2_{\mathbb{S}^3} - \frac{3N - 1}{2 a^2}
\left\{1 + \frac{3N - 2}{\kappa a}\right\}
\right\} 
\Psi + 2\kappa a\{\Lambda a^2 - 3\} \Psi + V(\mbox{3--}\sphericalangle, a) \Psi = 0  \mbox{ } . 
\ee
(This makes use of the conformal operator ordering \cite{Conf}.)
This would be solved to zeroth order in terms of $a$ and then subjected to inhomogeneous RPM particle perturbations.  

\vspace{10in}

\ni{\bf Acknowledgements} To those close to me gave me the spirit to do this.  
And with thanks to those who hosted me and paid for the visits: Jeremy Butterfield, John Barrow and the Foundational Questions Institute.
Thanks also to Chris Isham and Julian Barbour for a number of useful discussions over the years.

%=====================================================BIBLIOGRAPHY========================================================================================================================

\end{document}